# Interplay between the mechanics of bacteriophage fibers and the strength of virus-host links

P. Ares[1,2,3], C. García-Doval[4], A. Llauró[2,3], J. Gómez-Herrero[2,3], M. J. van Raaij[4] and P. J. de Pablo[2,3*]

[1]Nanotec Electrónica S.L., Ronda de Poniente, 12 – 2 C, 28760 Tres Cantos, Madrid, Spain
[2]Departamento de Física de la Materia Condensada, Universidad Autónoma de Madrid. 28049 Madrid, Spain.
[3]IFIMAC, Centro de Investigación de Física de la Materia Condensada. 28049 Madrid, Spain
[4]Departamento de Estructuras de Macromoléculas, Centro Nacional de Biotecnología (CNB-CSIC), calle Darwin 3, 28049 Madrid, Spain

Viral fibers play a central role in many virus infection mechanisms since they recognize the corresponding host and establish a mechanical link to its surface. Specifically, bacteriophages have to anchor to bacteria through the fibers surrounding the tail before starting the viral DNA translocation into the host. The protein gene product (gp) 37 from bacteriophage T4 long tail fibers forms a fibrous parallel homotrimer located at the distal end of the long tail fibers. Biochemical data indicate that, at least three of these fibers are required for initial host cell interaction, but do not reveal why three and no other number. By using Atomic Force Microscopy we obtained high-resolution images of gp37 fibers adsorbed on mica substrate in buffer conditions and probed their local mechanical properties. Our experiments of radial indentation at the nanometer scale provided a radial stiffness of ~0.08 N/m and a breaking force of ~120 pN. In addition, we performed Finite Element Analysis and determined a Young's modulus of ~20 MPa. From these mechanical parameters, we hypothesize that three viral fibers provide enough mechanical strength to prevent a T4 virus from being detached from the bacteria by the viral particle Brownian motion, delivering a biophysical justification of the previous biochemical data.

[*]p.j.depablo@uam.es

**Introduction**

Mechanical properties of biological molecular aggregates are essential to their function. Indeed, forces at the nanoscale play a central role in biochemistry, from the myosin-actin system [1], which is the ultimate responsible of muscle motion, to the DNA-related motor proteins [2]. Viruses are striking examples of biomolecular aggregates where recent studies of their individual mechanical properties have provided interesting insights into their physical functionality. Among others, these studies have unveiled the structural role that the DNA may play either reinforcing the shell or exerting pressure on the viral walls, [3-6], the influence of defects on their stiffness [7], the existence of mechanical pre-stress [8, 9], or the dependence of the mechanical properties on the conformational changes required for the infectivity [10, 11].

Together with the nucleic-acid containing capsid, a virus possesses other structures, such as the tails and the fibers, whose physical properties have not been studied yet with single molecule techniques despite their importance in the viral cycle. Specifically, viral fibers are present in many eukaryotic viruses and bacteriophages and they are responsible for the initial stages of infection [12]. For instance, human adenovirus fiber binds the coxsackievirus-adenovirus receptor protein, which is on the cell surface [13]. Many of the bacteriophages belonging to the *Caudovirales* order also use fiber proteins for host recognition and adhesion to the bacterial cell wall. In particular, bacteriophage T4, a *Myovirus*, has been studied extensively as a model system for assembly of complex structures [14]. In the case of phage T4, the initial recognition of the bacterial cell required for the infection, is carried out by the long tail fibers (depicted in Fig. 1a). These fibers reversibly bind to the outer glucose[α1-3]glucose region of the bacterial lipo-polysaccharide (LPS) or the Outer Membrane Porin C (OmpC) [15, 16]. Upon receipt of the signal —achieved when at least three long tail fibers have encountered suitable receptors— a conformational change of the baseplate [17] allows the short tail fibers, which are trimers of gene product (gp) 12, to extend [18]. Once these short tail fibers have irreversibly bound the core region of the LPS [19], a conformational change likely allows the inner tail tube to pass through the baseplate (an action driven by the contraction of the outer tail sheath). Phage proteins and DNA can then enter the bacteria and initiate infection which, in favorable conditions, can lead to several hundred daughter phages and bacterial lysis within 30 minutes [20].

The long tail fiber can be divided in proximal and distal halves, each over 70 nm long and connected at an angle of about 160º [21]. The half proximal to the phage (the thigh) is made up of a trimer of gp34, a 1289 amino acid protein of unknown structure. At the kink, a single copy of gp35 (372 residues) is located. The top of the shin is constructed of a trimer of the 221-amino acid protein gp36, while the major part of the shin and the receptor-binding tip (or foot) is comprised of a parallel homo-trimer of gp37. Long tail fiber structure is shown in Fig.1b. Full-length gp37 contains 1026 residues. The crystal structure of a trimer containing residues 811-1026 for each of the three chains has been resolved at 2.2 Å resolution [22]. The structure revealed a collar domain similar to that observed for gp12, [23] which is composed of amino acids 811-861 plus a β-strand formed by the very C-terminus of the protein (residues 1016-1026). This means the N- and C-termini of this fragment are close, and the intervening residues form an extensively interwoven and intertwined region (Fig. 1b) (residues 862-880 plus 1009-1015), a needle domain consisting of amino acids 881- 933 plus 960-1008 and a small head domain formed by residues 934-959. The head domain is responsible for initial host recognition. Since at least three fibers have to hold the capsid onto the bacterial cell wall, their mechanical strength must at least overcome the mechanical tension provided by the thermal Brownian motion of the virus. We report herein on the mechanical properties of gp37 fibers such as the Young's modulus, the breaking force and stiffness, which we have obtained by Atomic Force Microscopy and Finite Element Analysis (FEA). We relate these data with the mechanical function of the fibers during the first stage of viral infection, providing a hypothesis of why at least three fibers are needed to initiate T4 infection.

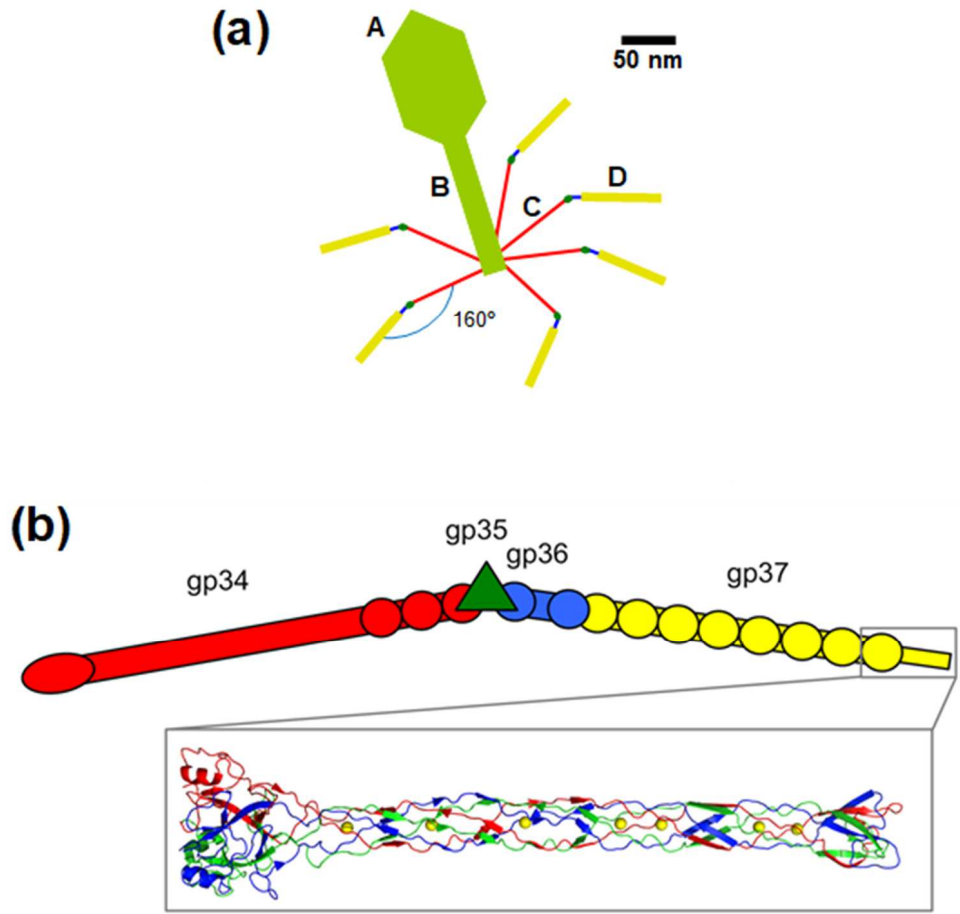

**Figure 1. Fibers structure.** (a) shows the relative position of gp37 fibers -yellow (thicker) lines- in phage T4: *A* virus capsid. *B* inner tail. *C* proximal half fibers. *D* distal half fibers. *C+D* long tail fibers. (b) shows the long tail fibers structure and protein domains. Detail in the rectangle shows the known crystallographic structure of the gp37 fiber.

**Materials and Methods (M&M)**

**A. Sample preparation**

For purification of amino-terminally six-histidine tagged gp37 containing residues 12-1026, cultures of JM109(DE3) transformed with pCDF (Sm)gp37 and pET(Ap)57 were grown at 37ºC in Luria Broth (LB) medium supplemented with ampicillin (100 mg/l) and streptomycin (50 mg/l) to an optical density of 0.6 units measured at 600 nm [24]. Expression was induced by adding isopropyl-β-D-thiogalactoside (IPTG) to a final concentration of 1 mM and growth was continued overnight at 16 ºC. Bacteria were harvested by centrifugation and pellets were resuspended in lysis buffer (50 mM Tris-HCl pH 8.0, 4 % (v/v) glycerol, 50 mM ammonium chloride, 2 mM EDTA and 150 mM sodium chloride). Cells were lysed by 10 rounds of 10 s sonications alternated with incubation on ice and extracts were centrifuged for 20 min at 20000 x *g* at 4ºC. Imidazole from a 1 M stock at pH 7.0 was added to the supernatants to give a final concentration of 50 mM and the resulting mixture was loaded onto a nickel-iminodiacetic acid agarose column equilibrated in buffer A (50 mM Tris-HCl pH 8.5; 300 mM sodium chloride) containing 50 mM imidazole. One ml resin slurry as supplied (Jena Bioscience, Jena, Germany) was used to form the column. Elution was done with six passes of 5 mL of buffer A containing 100, 150, 200, 250, 300 and 400 mM imidazole, respectively. The 150, 200 and 250 mM imidazole fractions were combined and dialysed against 10 mM Tris-HCl pH 8.5. The protein was applied to a 6 mL Resource Q column equilibrated with 10 mM Tris-HCl pH 8.5 and eluted with a sodium chloride gradient. Highly purified protein eluted at around 0.25 M sodium chloride and was dialysed against 10 mM Tris-HCl pH 8.5. A stock of purified gp37 fibers at 0.1 mg/ml in 10 mM Tris-HCl buffer at pH 8.5 was used.

**B. Atomic Force Microcopy**

For AFM experiments, the stock solution was diluted 20 times to 5 µg/ml. To attach the fibers to the substrates, surfaces of freshly cleaved mica discs were immersed in a solution of 0.1% of 3-aminopropyl-triethoxy-silane (APTES) (Sigma–Aldrich). Then the mica discs were rinsed with 2-propanol and water and dried in a $N_2$ gas jet. A 40 µl drop of diluted stock was deposited onto a treated mica disc. The drop was left on the surface for 20 min and, afterwards, rinsed four times with 40 µl drop of buffer without allowing surface dewetting. The tip was also pre-wetted with 30 µl drop of buffer solution in order to perform the experiments in liquid. A Cervantes Fullmode AFM (Nanotec Electrónica, Madrid, Spain, www.nanotec.es) was operated in buffer at room temperature using non contact Dynamic mode [25] through the WSxM software [26]. A cantilever holder specially designed to work in this mode in liquid environments was used [27] to guarantee an optimum operation of the microscope. We used silicon nitride rectangular cantilevers (OMCL-AC40TS; Olympus, Tokyo, Japan, http://probe.olympus-global.com). The spring constants of the four different cantilevers used for the measurements were calibrated by using Sader's method in air [28] yielding 0.06 ± 0.01 N/m, with average tip radius of 11 nm [29] (Supplementary Information (SI), [30] Fig. S1) and resonance frequency values in buffer conditions of ~20 kHz. Low spring constant cantilevers were chosen due to their high force sensitivity, which is convenient for soft samples.

The cantilevers were oscillated with amplitudes of ~3 nm at their resonance frequency. In order to minimize the tip-sample interaction and increase the scan rate, images were only acquired while the tip was scanning from left to right (trace). The retrace was acquired much faster but with the tip far away from the surface. AFM images and spectroscopy data were processed by using WSxM software. After locating individual fibers on the surface, the selected sample was zoomed in, and the scan was performed always on the same line, with the fiber filling near the whole image range. Then the lateral piezo scan was stopped when the tip was on top of the fiber. At this point, a force *vs*. z-piezo position (FZ) curve was performed by elongating the *z*-piezo so that the tip established mechanical contact with the fiber (nanoindentation). After each FZ measurement the fiber was immediately scanned to check for structural integrity. Thus, about two or three consecutive nanoindentation-imaging cycles were performed on the fiber until a breakage event was observed in both indentation curve and AFM topography.

**C. Young's modulus calculations with FEA**

To calculate the Young's modulus values from FZ curves, these data had to be converted into force *vs*. indentation F($\delta$) curves. This standard process consist on comparing the measured FZ curve on the sample with the FZ curve on the substrate (mica in our case). At any applied force value, the corresponding difference in z-distance between the FZ curves provides a measure of the fibers indentation at that force value [31]. The resulting indentation curve can be fitted to physics-based models that predict the AFM tip-sample contact mechanics, and the Young's modulus can be estimated by tuning the theoretical Young's modulus value to match the theoretical prediction with the experimental data [32, 33].

FEA was performed using COMSOL Multiphysics$^{TM}$ 4.1. The viral fiber was made of a homogenous material (SI, [30] Fig. S2) with a Poisson ratio of 0.3, according to studies with similar conditions [9, 34, 35] and taking into account that previous FEA on viral capsids showed that the models were quite insensitive to the Poisson ratio variation [36]. The viral fiber was modeled mimicking the dimensions and geometry of the fiber structure obtained from Electron Microscopy (EM) volume data [21]. The fiber was placed on a flat rigid surface and was loaded with a rigid 11 nm radius indenter, similar to the AFM tips used in our experiments. The contacts between the fiber and the tip and between the fiber and the surface were both implemented with a contact-penalty stiffness method according to the Comsol manual. Considering the symmetry of the geometry, the model was reduced to one half and meshed over 6801 tetrahedral elements. A parametric, non-linear solver was used to simulate the stepwise lowering of the indenter onto the fiber. To match the experimental curves with the simulation, the Young's modulus was varied until accordance between simulation and experiment was successfully reached.

**D. Brownian force of phage T4**

Assuming that the shape of the T4 can be approximated by a sphere of radius r (~50 nm), its dragging coefficient is given by the Stoke's Law as $\gamma = 6\pi\eta r = 9.4\times10^{-10}$ Nsm$^{-1}$, where $\eta$ is the water viscosity (0.001 Pa s) [37, 38]. As a consequence of the Principle of Equipartition, the root mean square (RMS) virus velocity at room temperature T is given by $\langle v^2 \rangle^{1/2} = \left(\frac{3k_B T}{m}\right)^{1/2} = 0.20 \, m/s$, where $k_B$ is the Boltzman constant and m the mass of the phage T4 (~190 MDa) [39]. The persistence time $\tau$, that represents the lag of time in which the virus moves in a given direction due to the thermal power stroke is given by $\tau = \frac{m}{\gamma} = 335 ps$. On average, during this time the virus particle moves a distance $d = v \times \tau = 0.7$ Å in a random direction, which represents about ~1/1400 of the phage diameter. The variance of the power spectrum of the thermal force of a tethered virus is independent of the frequency and only depends on the drag coefficient, and not on the stiffness of the tether, i.e. $\langle F^2 \rangle = \gamma^2 \langle v^2 \rangle$ [40]. From this formula and the standard deviation of the velocity provided by the Maxwell-Boltzmann distribution of velocities, i.e. $\sigma = \sqrt{\left(\frac{3\pi-8}{\pi}\right)\left(\frac{k_B T}{m}\right)}$ we derive the force for every thermal shake on the viral particle to be ~190 ± 70 pN.

**Results**

After fibers have anchored to the substrate, dynamic AFM imaging of the surface in buffer conditions reveals a random dispersion of elongated structures (Fig.2). Each of them shows several longitudinal bumps which correspond fairly well to the different protein modules composing the gp37 viral fibers (Fig.3a) [22].

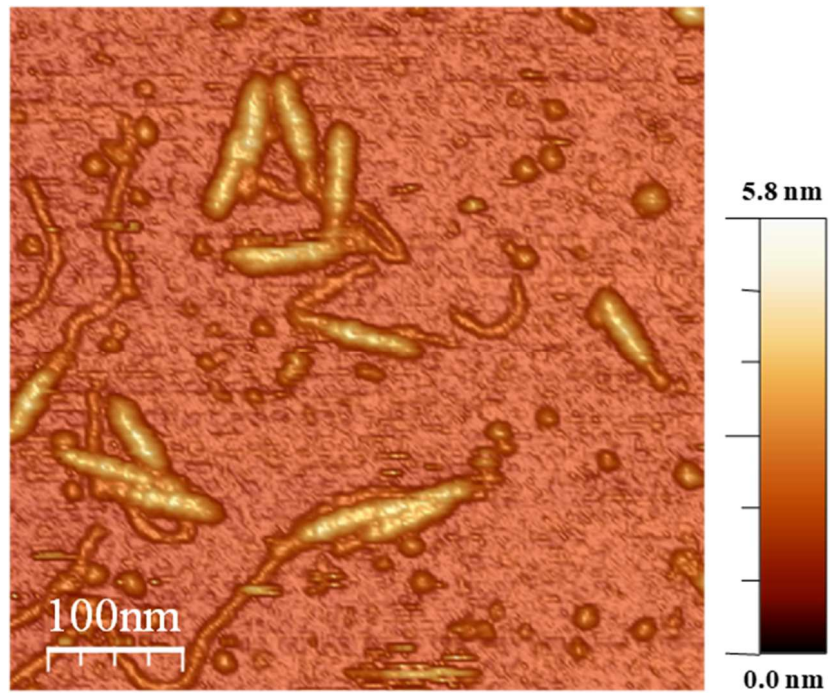

**Figure 2. gp37 fibers on mica.** This topographical AFM image represents a random population of gp37 fibers on mica under buffer conditions.

An inherent effect of AFM imaging of samples with a similar dimension to the tip radius is the increase of lateral size, although the height remains unchanged. Thus, we can analyze the topography of the fibers by calculating the expected geometrical dilation effect between the tip and the proteins. Models of the gp37 fibers were created based on the EM volume of the fibers [21]. Therefore we processed the fibers' EM data (Fig. 3a) with a dilation algorithm using a parabolic tip ($z = x^2/2r$) [41]. For the radius r we chose 11 nm. The resulting dilated fiber (Fig. 3b) presents topographical features in good agreement with the AFM image of Fig. 3c. Processed data reveals that fiber width is highly affected by dilation. Conversely, since the fiber is much longer than the tip radius, its length remains unchanged by the lateral dilation. We have measured the height and the length of 55 viral fibers, obtaining an average height of 4.7 ± 0.2 nm (5.1 ± 0.3 nm and 4.3 ± 0.3 nm for the protein moduli and inter-protein moduli loci, respectively) and a length of 64 ± 5 nm (SI, [30] Fig. S3), both in good agreement with the dimensions expected from the fiber proteins and from previous EM data (Fig. 3a) [21, 24]. The comparison of the height profiles along the fiber between EM and AFM data (Fig. 3d) also indicates a reasonable match, bearing in mind that the EM volume is the result of averaging images of many fibers while the AFM volume is the result of scanning only one fiber.

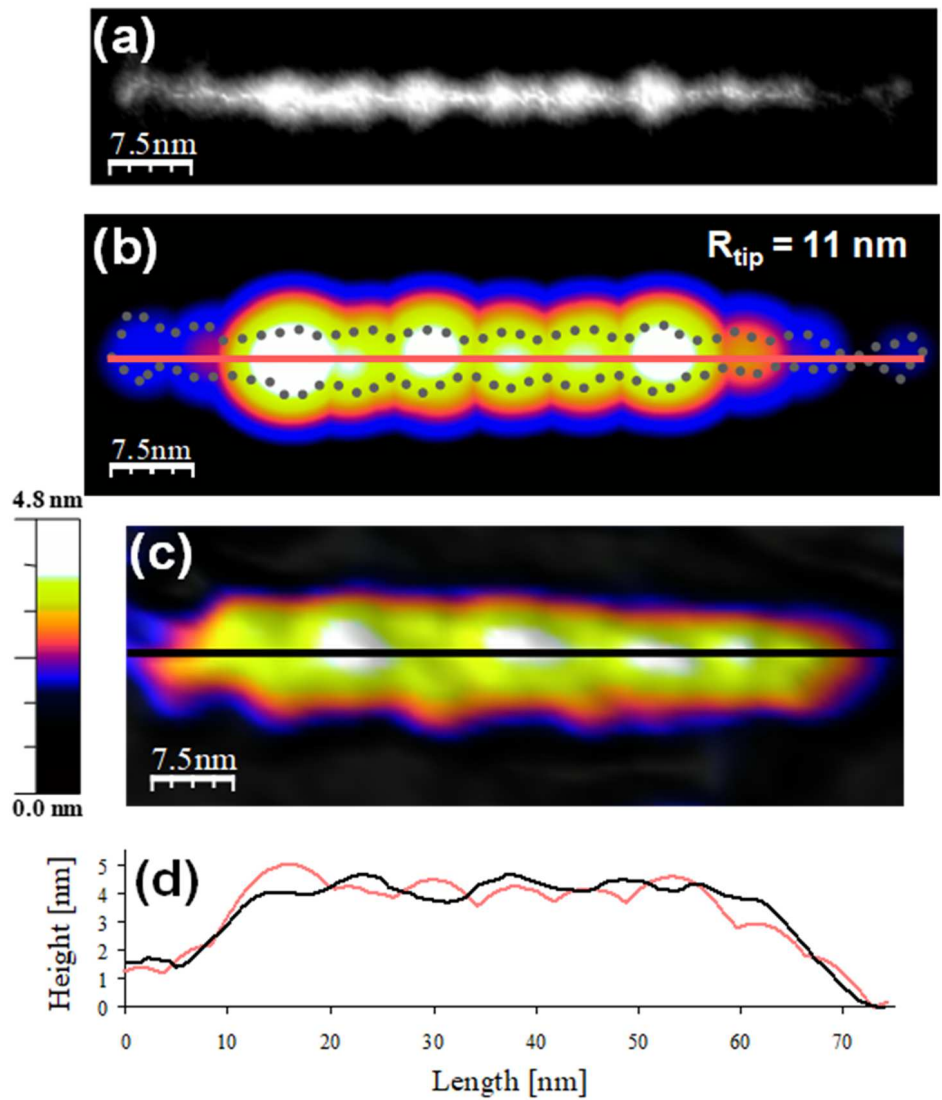

**Figure 3. Topographic analysis and EM-AFM data comparison.** (a) presents EM volume data of gp37 [21]. (b) shows a geometrical dilation filtering of EM data with a tip of radius 11 nm. The dotted contour highlights the EM data for the sake of comparison. (c) presents experimental topographical AFM data of a gp37 fiber adsorbed on mica. Horizontal profiles in (b) red (light gray) and (c) black, are depicted in (d).

For the indentation experiments, once a single fiber is selected on the surface, we perform a nano-indentation on its top. Afterwards, an image of the fiber is acquired to check its integrity (M&M). The procedure is repeated until the fiber is broken. Figure 4 presents the AFM image of a fiber before (Fig. 4a) and after (Fig. 4b) the dismantling provoked by the FZ of Fig. 4c. Figure 4b demonstrates the dramatic modification caused by the nanoindentation curve, as we observe that several protein moduli have been ripped out by the tip. The mechanism of the fiber indentation (Fig. 4c) is performed as follows: after taking a reference FZ on mica (red), the cantilever approaches to the fiber at zero deflection until it establishes mechanical contact. Afterwards, the tip starts deforming the fiber, showing a linear deformation from point A to point B, where a steep drop of the cantilever indicates that the fiber is broken and the tip taps the mica surface.

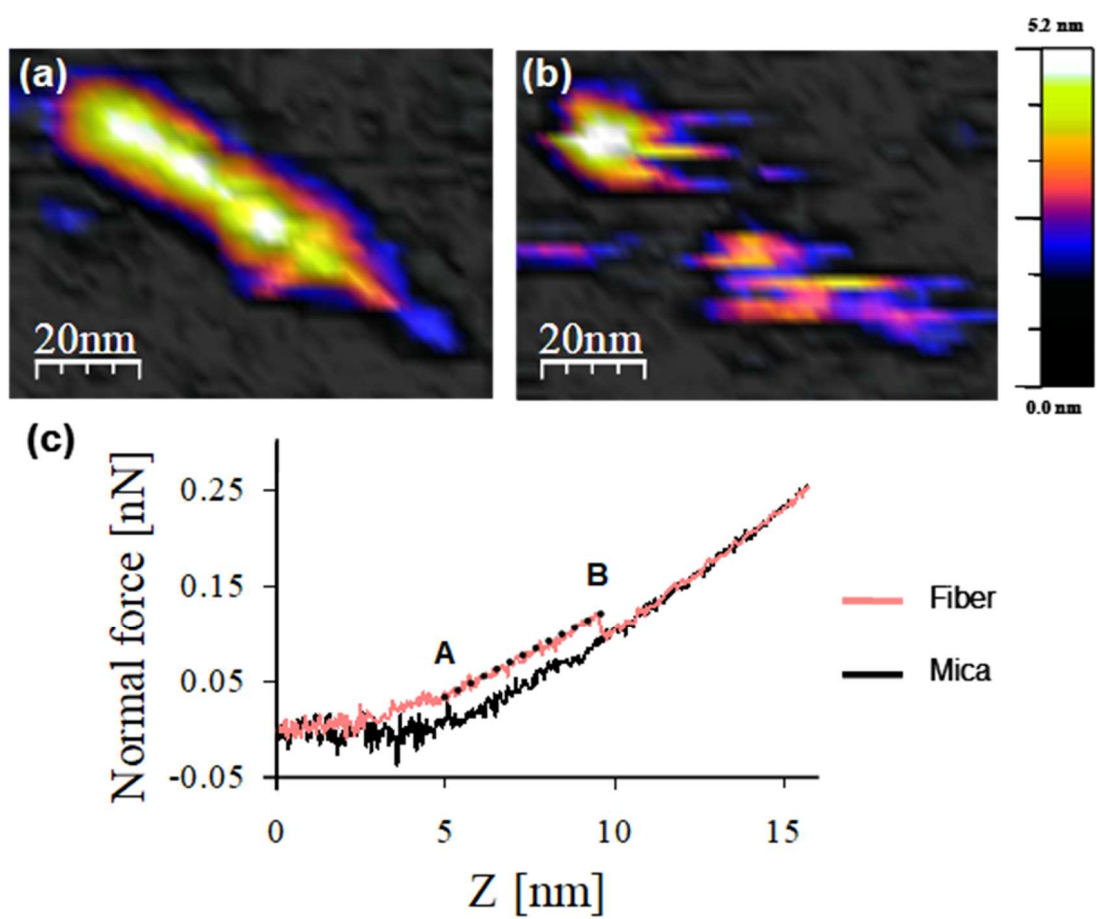

**Figure 4. Breaking a gp37 fiber.** (a) and (b) present the topographical images of a fiber before and after a nanoindentation experiment. Forward cycles of nanoindentation on the fiber and on the mica substrate are depicted in (c) red (light gray) and black, respectively. Point *A:* marks the starting point of the linear deformation of the fiber. Dotted line depicts the linear fit for extracting the stiffness of the fiber. Point *B*: the fiber is broken.

Figure 5a presents 18 indentation curves performed on 18 different fibers, where the surface deflection signal has been previously subtracted from the fiber deflection one in order to obtain the deformation of the fiber [31, 42]. To this end, it is assumed that while the tip can squeeze the fiber, it cannot deform the surface of the substrate (the substrate stiffness is much greater than the cantilever force constant). The contact point between the tip and the fiber is determined by the change in the force slope and the subsequent force noise reduction. While the tip approached the fiber the cantilever was free and no force was measured, but when the tip reached the top of the fiber indentation started, the slope of the curves changed abruptly and the force noise decreased [43]. By shifting the indentation curves to coincide with the tip-fiber contact points of all the curves, we can observe a variety in the distance indented until the substrate is reached by the tip. Most of the indentation ranges from about 4 to 5 nm, in good agreement with FZs performed at inter-protein modulus loci and at top of a protein modulus (SI, [30], Fig. S3).

Figure 5b-c show FEA simulations corresponding to the tip indenting at inter-protein modulus and at the top of protein modulus loci.

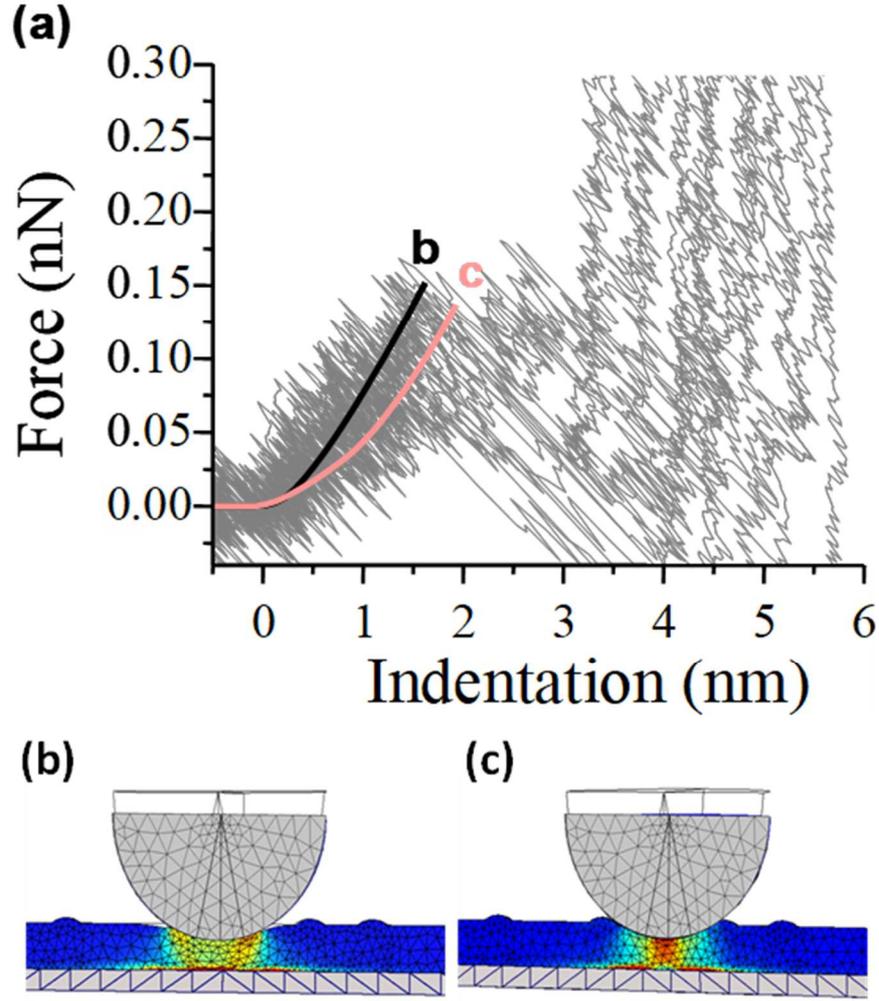

**Figure 5. Indentation curves on gp37 and FEA.** (a) presents 18 indentation graphs pertaining to 18 different fibers and the data fits from the FEA: black for the tip indenting on a protein moduli (b) and red (light gray) for the tip indenting on an inter-protein moduli (c).

## Discussion

It is instructive to estimate the approximate viral fiber stiffness by considering both cantilever and fiber as two springs in series with spring constants $k_{cl}$ and $k_{fib}$, respectively, thus obtaining [6]:

$$\frac{1}{k_{ms}} = \frac{1}{k_{cl}} + \frac{1}{k_{fib}} \qquad (1)$$

where $k_{ms}$ is the spring constant measured from point A to point B in a nanoindentation event (Fig. 4c) that accounts for simultaneous deformations of the cantilever and the fiber.

Assuming a linear deformation, we estimate an average gp37 fibers radial stiffness of $k_{fib}$ = 0.08 ± 0.03 N/m, similar to other bacteriophage bodies such as phi29 [9].

To perform a more profound mechanical characterization of the fibers, Finite Element Analysis (FEA) of the tip-fiber system has been done (M&M)) to estimate the Young's modulus of the fibers. Figure 5a shows the experimental data corresponding to indentations randomly distributed along 18 fibers and the FEA fits corresponding to two limit cases: the tip indenting on the top of a protein modulus (Fig. 5b) and on an inter-protein modulus (Fig. 5c). The comparison between the indentation experiments and the afore mentioned cases of the simulation results in a value of the Young's modulus of the fiber $E_{fib}$ = 20 ± 5 MPa.

Using the obtained Young's modulus we can calculate the longitudinal stiffness of the fibers. From the Hooke's law the stiffness of an isotropic cylinder along its main axis can be expressed as follows:

$$k_{fib} = \frac{E_{fib} A_0}{L_0} \quad (2)$$

where $A_0$ is the cross-section through which the force is applied and $L_0$ is the length of the object. Figure 1b illustrates that the homo-trimer forms an extensively interwoven intertwined region where the three proteins are heavily coupled. Therefore, in first approximation, as it is commonly done in such kind of analysis, we can assume the fibers as isotropic cylinders [33, 44] (SI, [30] Fig. S2), where $A_0$ will be the cross-sectional area of the fibers and $L_0$ their length. Thus, the longitudinal stiffness results in $k_{long}$ = 0.005 ± 0.002 N/m.

The experiments also allow us to extract other important mechanical parameters such as the radial breaking force $F_{rad}$ = 120 ± 30 pN and a collapse distance of 2.0 ± 0.3 nm (SI, [30] Fig. S4). From both the FEA and Hertz model (SI, [30]), we can extract the average area of contact between the tip and the fiber at the radial breaking force to be s = 13.5 ± 1.5 nm$^2$, and derive a tensile strength of $\sigma_r = \frac{F_{rad}}{s} = 9 \text{ MPa}$. The fiber cross-section $A_0$ is calculated by using its semi-height (see Results) as the average radius, to give 17 nm$^2$. Thus assuming isotropy, we can estimate the longitudinal breaking force $F_{long} = \sigma_r \times A_0$ = 150 ± 30 pN.

To our knowledge, this is the first measurement of the mechanical properties on viral fibers, hindering the comparison with previous reports. Nevertheless, we can compare our results with other protein fibers [45]. In particular, fibrin fibers, which are the major structural component of a blood clot, are extraordinarily extensible and elastic, and they are relatively soft [46-48]. Although fibrin fibers lengths and diameters are much larger than gp37 fibers, fibrin fibers are assembled from fibrin monomers, which come from the removal of two pairs of fibrinopeptides in the fibrinogen. Fibrinogen is a highly abundant, soluble plasma protein and it is 46 nm in length and 4.5 nm in diameter [49], which are similar to gp37 fibers dimensions. The Young's modulus and the rupture force per monomer of fibrin fibers is 1-10 MPa [46, 48, 50], and 280 pN [47], respectively. Since the Young's modulus is a bulk material property, it is not surprising that fibrin fibers show similar values to gp37 fibers because the ultimate building blocks of both structures are individual proteins.

On the other hand, it is convenient to discuss the biophysical implications of our results. T4 phage infection is initiated by the attachment of at least three long tail fibers to the host [51]. This process triggers a conformational change of the base plate that induces the tail expansion and the subsequent DNA translocation into the host [38]. During this process long tail fibers have to withstand detaching forces, such as those caused by the Brownian fluctuations, the mechanical tensions derived from the multiple conformational changes occurring during infection, or the exchange of momentum provided by the DNA translocation. In particular, the mechanical resistance of the first three bound fibers against the thermal shaking of the viral particle is critical for infectivity. The thermally activated Brownian motion of a viral particle consists on random displacements of a distance $d \sim 0.7$ Å every 300 ps on average (M&M)). When the virus is pulled normally from the bacteria surface by thermal vibrations, both fiber and fiber-receptor are subjected to an average force $F_{thermal} = 190 \pm 70$ pN (M & M). We can compare $F_{thermal}$ with the longitudinal breaking force of a fiber, $F_{long}$ to hypothesize about why at least three fibers are required for initiating T4 infection. By accounting the phage fibers as parallel tethers anchoring the viral particle to the bacteria wall, we can estimate the breaking force for 1, 2 and 3 fibers (Fig. 6). We find that 1 and 2 fibers are insufficient to hold the virus particle on the bacteria, and only three or more fibers provide enough resistance to the thermal force.

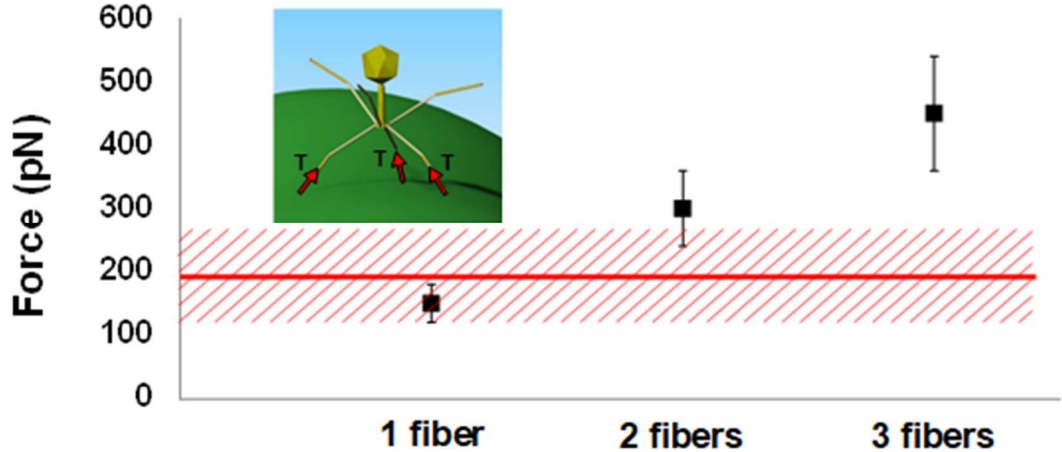

**Figure 6. Thermal force and fibers breaking forces.** The horizontal red line depicts the thermal force $F_{thermal}$ of T4 bacteriophage, 190 ±70 pN. Squares represent the forces needed to break 1, 2 and 3 fibers. Inset: cartoon of a T4 virus attached to a bacterium through three of its long tail fibers, showing the tension on each of them during a thermal power stroke.

After recognition has occurred and the virus is bound to the bacteria, a variety of conformational changes allows the short tail fibers to bind to the outer region of the bacterial lipo-polysaccharide and the inner tail tube to pass through the baseplate. As a consequence of these conformational changes the long and short tail fibers, together with the tube of the tail hold the virus attached to the bacteria and the DNA translocation starts, exerting a force of ~60 pN [52]. The longitudinal breaking force of an individual long tail fiber (~150 pN) is about twice the force induced by DNA translocation. Thus, the three fibers do not only withstand the force induced by thermal shaking during the fiber-host recognition stage, but also withstand the forces during the rest of the infection process.

**Conclusions**

We have obtained high-resolution AFM images of the bacteriophage T4 gp37 fibers under close to physiological conditions. By using radial indentations together with Finite Element Analysis, we have determined their Young's Modulus (20 MPa), their radial stiffness (0.08 N/m), and their radial and longitudinal breaking forces. With these results we estimated the mechanical resistance of the fibers to hold the viral particle on the bacteria of being detached by Brownian motion. Thus, we hypothesized that at three fibers provide enough mechanical resistance for initiating the infection. These results and further studies, such as for example the use of bacterial lipo-polysaccharide and isolated Outer Membrane Porin C, or the performance of pulling experiments, could lead to a better understanding of the mechanical basis of viral-host recognition and infection mechanisms.

**Acknowledgements**

We acknowledge funding from grants MAT2008-02533, PIB2010US-00233, FIS2011-29493 and FIS2011-16090-E (P.J.P.), BFU2008-01588, BFU2011-24843 and BIO2011-14756-E (M.J.v.R) from the Spanish Ministry of Economy and Competitiveness, an FPU Ph.D. fellowship from the Ministry of Education (C.G.D.) and the Red Española Interdisciplinar de Biofísica de los Virus (BioFiViNet). We also thank A. Gil and M. Hernando-Pérez for insightful discussions and F.J. García-Vidal for the help with Finite Element Analysis.

**Supplementary Information**

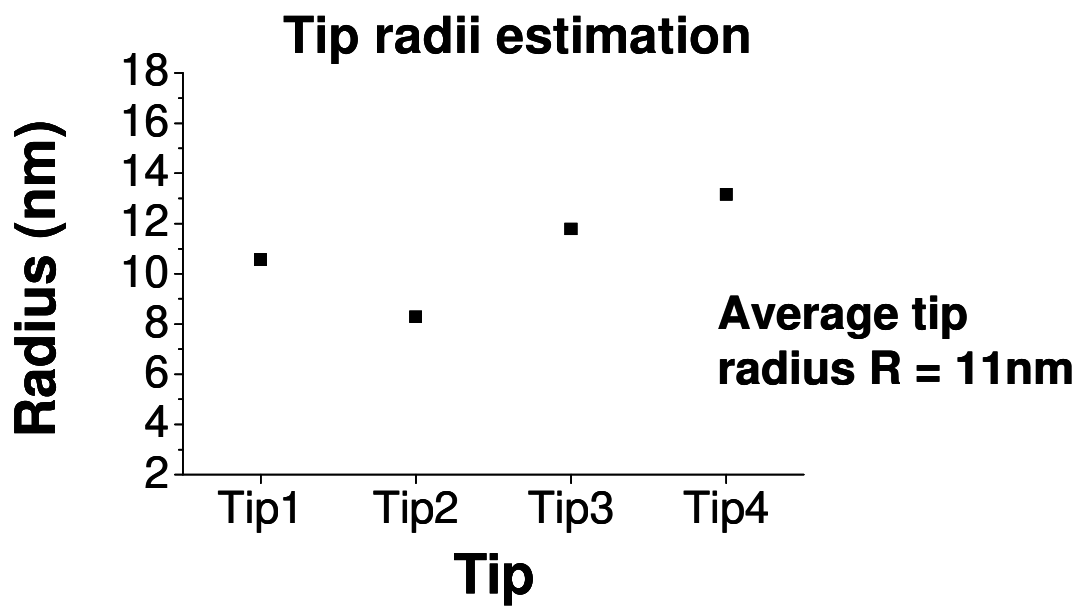

**Figure S1. Tip radii values for the four different tips used in the measurements.** The tip radius has been estimated from the topographic images of the fibers using the expression: $R = \dfrac{w^2}{8h}$, where $R$ is the tip radius and $w$ is the apparent width measured for a fiber of height $h$ (reference [29] in the article [P. Markiewicz and M. C. Goh, Langmuir **10**, 5 (1994)]).

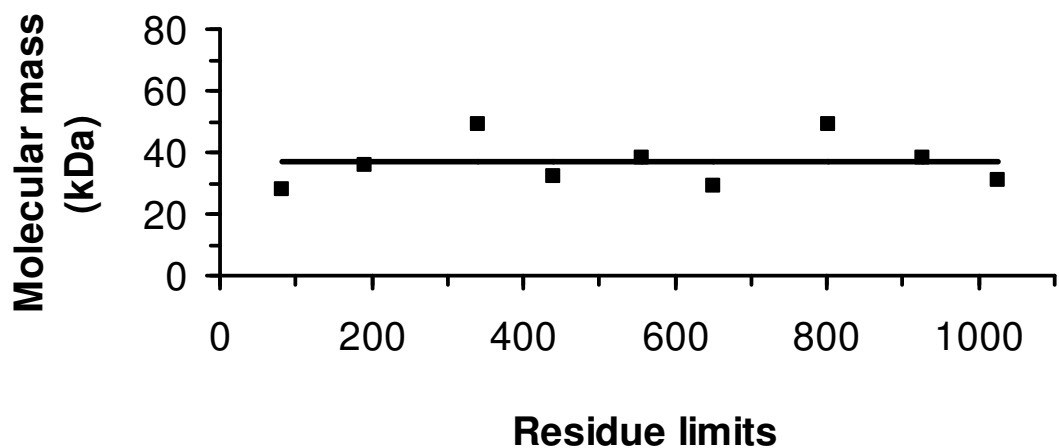

**Figure S2. Fiber homogeneity.** Molecular masses as a function of the residue limits in gp37 distal half-fibers. Data taken from reference [21] in the article [M. E. Cerritelli, J. S. Wall, M. N. Simon, J. F. Conway, and A. C. Steven, Journal of Molecular Biology **260**, 767 (1996)]. Since the dispersion of the values is low, in a first approximation the fibers can be considered homogeneous. The horizontal line shows the average of the molecular mass along the fiber. This average corresponds to 37 ± 7 kDa.

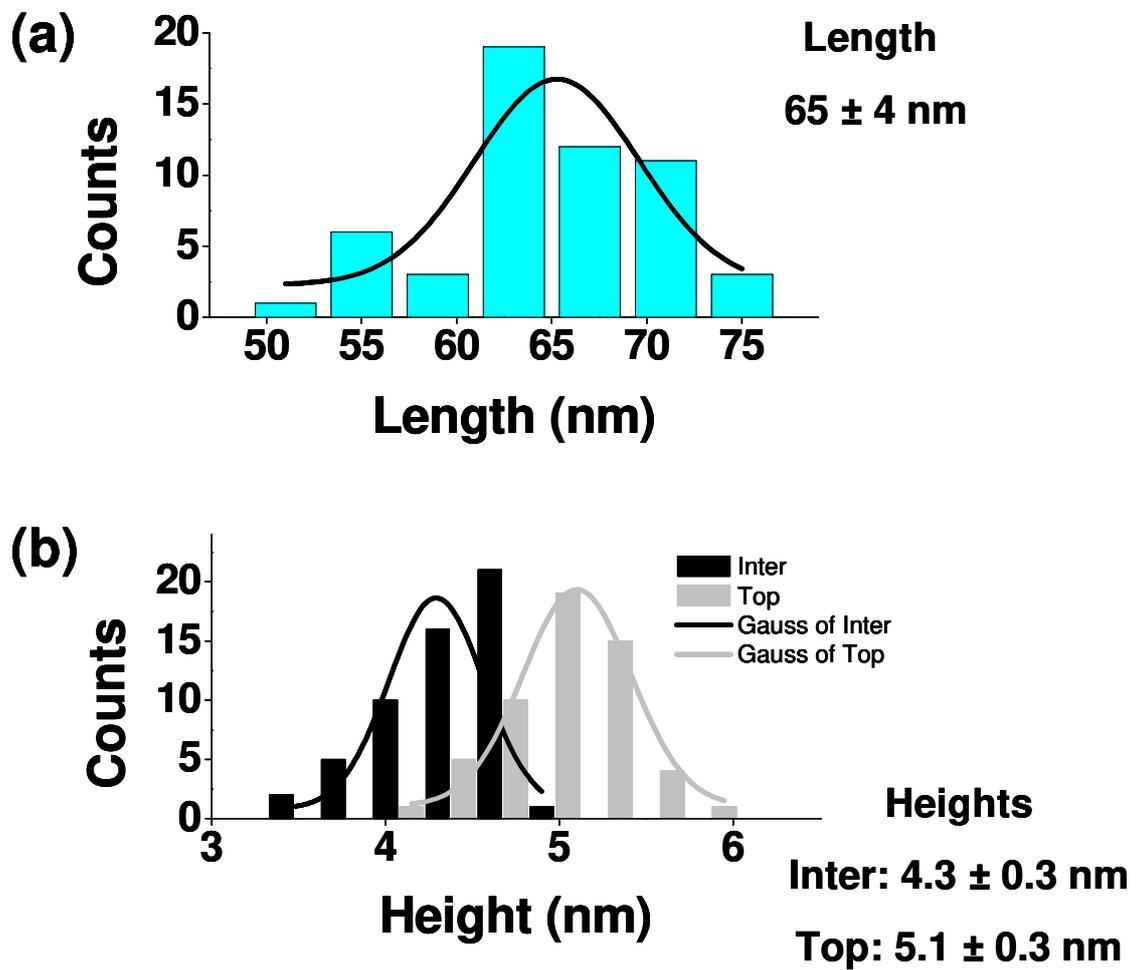

**Figure S3. Statistics of fiber dimensions.** (a) Length distribution. (b) Heights distribution. Inter and Top correspond to the heights at an inter-protein modulus and at the top of a modulus protein loci respectively. Average values have been obtained from Gaussian fitting (*mean ± SD*).

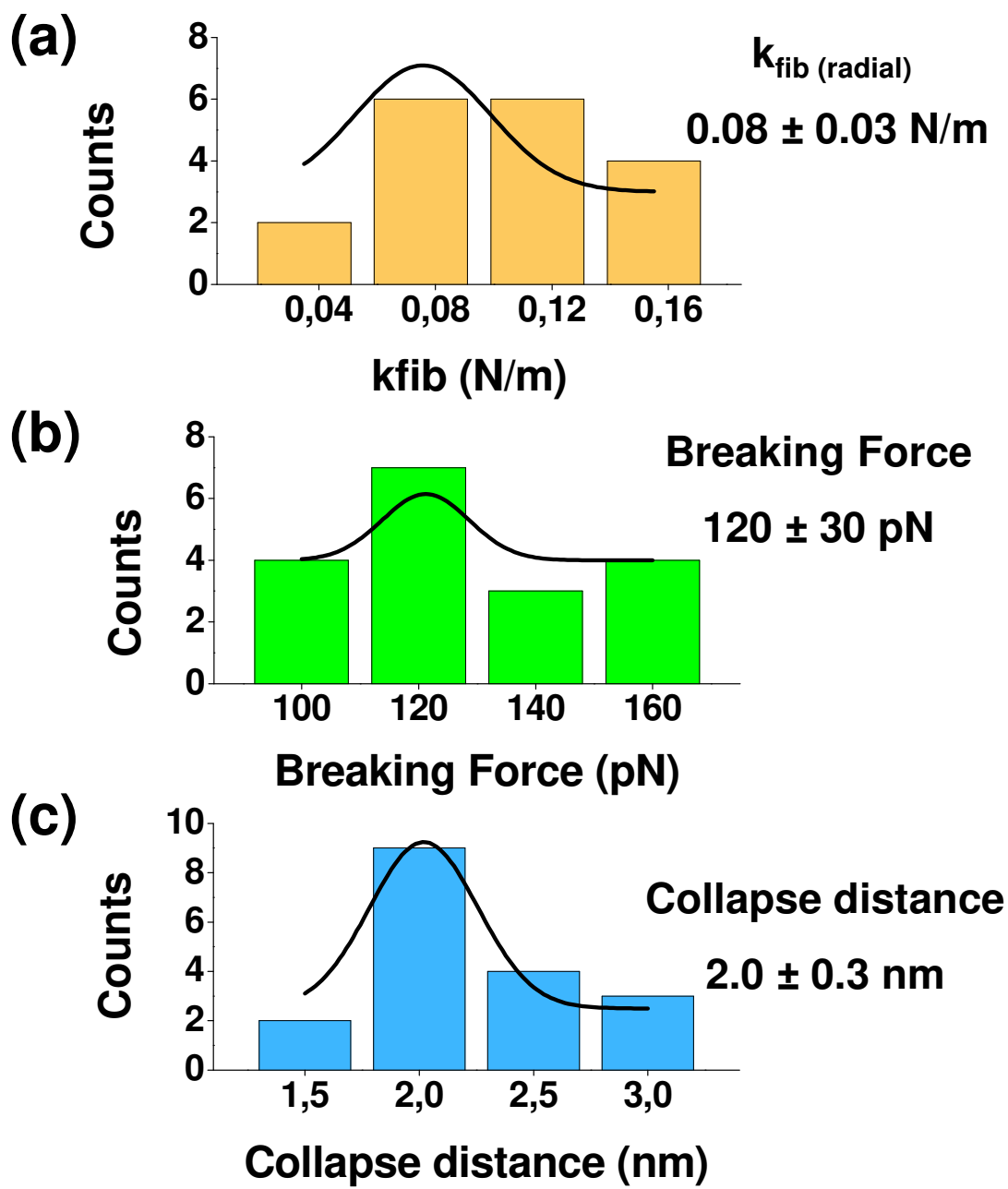

**Figure S4. Statistics of mechanical properties data.** (a) Radial stiffness distribution. (b) Breaking force distribution. (c) Fiber collapse distance distribution. Average values have been obtained from Gaussian fitting (*mean ± SD*).